\documentclass[useAMS,usenatbib]{mn2e}
\usepackage{epsfig}
\usepackage{epstopdf}
\usepackage{lscape} 
\usepackage{graphicx}
\usepackage{multirow}
\usepackage{hhline}
\usepackage[fleqn]{amsmath}
\usepackage{gensymb}
\usepackage{commath}
\usepackage{float}
\usepackage[flushleft]{threeparttable}

\begin{document}

\title[Lensing geometry]{
Pulsar lensing geometry
}

\author[Liu et al]{Siqi Liu$^{1,3}$\thanks{E-mail:\ sqliu@cita.utoronto.ca}, Ue-Li
  Pen$^{1,2}$\thanks{E-mail:\ pen@cita.utoronto.ca}, J-P Macquart$^{4}$\thanks{E-mail:\ J.Macquart@curtin.edu.au},
  Walter Brisken$^{5}$\thanks{Email:\ wbrisken@aoc.nrao.edu}, Adam Deller$^{6}$\thanks{E-mail:\ deller@astron.nl}\\
 $^1$ Canadian Institute for Theoretical Astrophysics, University of Toronto, M5S 3H8 Ontario, Canada \\
$^2$ Canadian Institute for Advanced Research, Program in Cosmology
and Gravitation\\
$^3$ Department of Astronomy and Astrophysics, University of Toronto, M5S 3H4, Ontario, Canada\\
$^4$ ICRAR-Curtin University of Technology, Department of Imaging and Applied Physics, GPO Box U1978, Perth, Western Australia 6102, USA \\
$^5$ National Radio Astronomy Observatory, P.O. Box O, Socorro, NM 87801, USA\\
$^6$ ASTRON, the Netherlands Institute for Radio Astronomy, Postbus 2, 7990 AA, Dwingeloo, The Netherlands\\
}

\date{\today}

\pagerange{\pageref{firstpage}--\pageref{lastpage}} 
\pubyear{2015}

\maketitle
\label{firstpage}
\begin{abstract}
Our analysis of archival VLBI data of PSR 0834+06 revealed 
that its scintillation properties can be precisely modelled using the inclined sheet model \citep{2014MNRAS.442.3338P}, 
resulting in two distinct lens planes.  
These data strongly favour the
grazing sheet model over turbulence as the primary source of
pulsar scattering.  
This model can reproduce the parameters of the observed diffractive
scintillation with an accuracy at the percent level.   
Comparison with new VLBI proper motion results in a
direct measure of the ionized ISM screen transverse velocity.  The results are consistent
with ISM velocities local to the PSR 0834+06 sight-line (through the Galaxy).
The simple 1-D structure of the lenses opens up
the possibility of using interstellar lenses as precision probes for
pulsar lens mapping, precision transverse motions in the ISM, and new
opportunities for removing scattering to 
improve pulsar timing.
We describe the parameters and observables of this double screen
system.  While relative screen distances can in principle be
accurately determined,
a global conformal distance degeneracy exists that allows a rescaling
of the absolute distance scale.  
For PSR B0834+06, we present VLBI astrometry results that provide (for the first time) a direct measurement of the distance of the pulsar.
For targets where independent distance measurements are not available, which are the cases for most
of the recycled millisecond pulsars that are the targets of precision timing observations, the degeneracy presented in the lens modelling could be broken if the pulsar resides in a binary system.

\end{abstract}
\begin{keywords}
Pulsars: individual (B0834+06) -- scattering -- waves -- magnetic: reconnection -- techniques: interferometric -- ISM: structure 
\end{keywords}

\newcommand{\be}{\begin{eqnarray}}
\newcommand{\ee}{\end{eqnarray}}
\newcommand{\beq}{\begin{equation}}
\newcommand{\eeq}{\end{equation}}

\section{Introduction}

Pulsars have long provided a rich source of astrophysical information
due to their compact emission and predictable timing. One of the
least well constrained parameters for most pulsars is their 
distance.  For some pulsars, timing parallax or VLBI parallax has
resulted in direct distance determination.  
For most pulsars, the
distance is a major uncertainty for precision timing interpretations,
including mass, moment of inertia \citep{2006Sci...314...97K,2012hpa..book.....L}, and
gravitational wave direction \citep{boyle2012}.

Direct VLBI observations of PSR B0834+06 show multiple images lensed
by the interstellar plasma.  
Combining the angular positions and
scintillation delays, the authors \citep{2010ApJ...708..232B} (hereafter B10) published the derived effective
distance (defined in Section \ref{21}) of $1168\pm 23$ pc
for apexes on the main scattering axis.
This represents a precise
measurement compared to all other attempts to derive distances to this
pulsar.  This effective distance is a combination of pulsar-screen and
earth-screen distances, and does not allow a separate determination of
the individual distances.  A binary pulsar system would in principle
allow a breaking of this degeneracy \citep{2014MNRAS.442.3338P}. One
potential limitation is the precision to which the lensing model can
be understood.  

In this paper, we examine the geometric nature of the lensing screens.
In B10, VLBI astrometric mapping directly demonstrated the highly
collinear nature of a single dominant lensing structure.  First hints
of single plane collinear dominated structure had been realized in
\citet{2001ApJ...549L..97S}.   While the nature of these structures
is already mysterious, for this pulsar, in particular, the puzzle is compounded by an offset group
of lenses whose radiation arrive delayed by 1 ms relative to the bulk of the pulsar flux.  The mysterious nature of
lensing questions any conclusions drawn from scintillometry as a
quantitative tool \citep{2014MNRAS.440L..36P}.

Using archival data we demonstrate in this paper that the lensing screen
consists of nearly parallel linear refractive structures, in two
screens.  The precise model confirms the one dimensional nature of the scattering geometry, and
thus the small number of parameters that
quantify the lensing screen. 

The paper is structured as follows. Section \ref{sec:lensing}
overviews the inclined sheet lensing model, and its application to data.
Section \ref{sec:astrometry} presents new VLBI proper motion and
distance measurements to this pulsar.
Section \ref{sec:discussions} describes the interpretation of the
lensing geometry and possible improvements on the observation.   
We conclude in Section \ref{sec:conclusions}.

\section{Lensing}
\label{sec:lensing}
\begin{figure*}
\centering
\includegraphics[width=1.0\linewidth, angle=0]{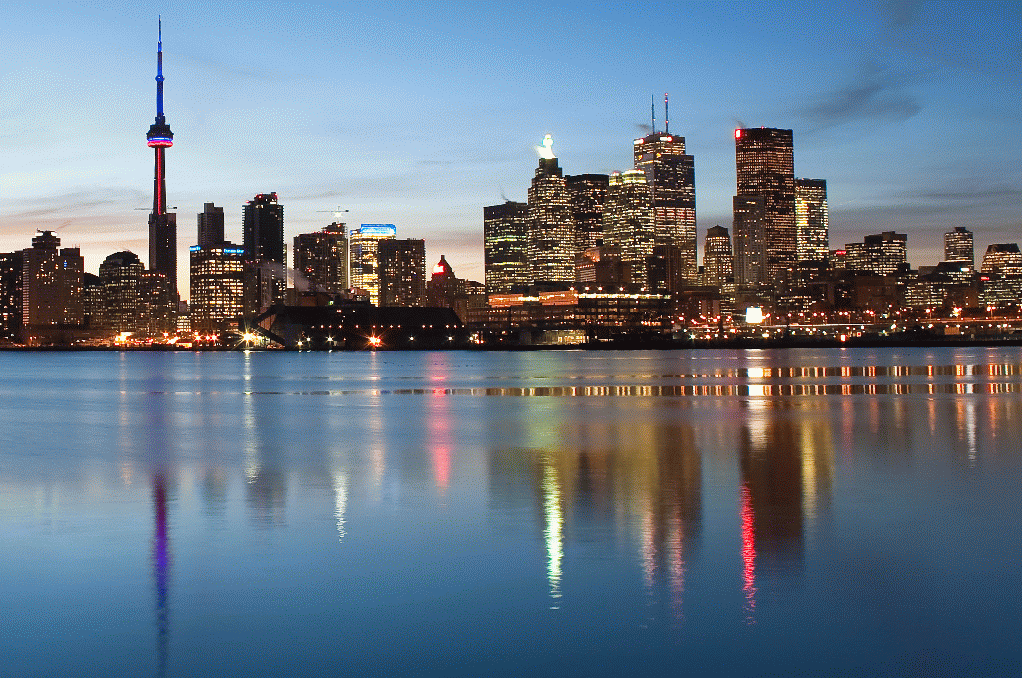}
\caption{Reflection of lights on surface waves.  At grazing angle,
 each wave crest results in an apparent image, causing a linear
 streak of images centred on the unperturbed image location.  For
 example, the red light streak would consist of a single image at its
 center in the absence of waves. The inclined sheet model for  pulsar
 scintillation is analogous, with reflection replaced by refraction.
Image copyright Kaitlyn McLachlan,
 licensed through shutterstock.com image ID 45186139.}
\label{fig:water_reflection}
\end{figure*}

In this section, we map the archival data of PSR 0834+06 onto the grazing incidence sheet
model.  The folded sheet model is qualitatively analogous to the
reflection of a light across a lake as seen from the opposite
shore.
In the absence of waves, exactly one image forms at the point
where the angle of incidence is equal to the angle of reflection.  In
the presence of waves, one generically sees a line of images above and
below the unperturbed image.  The grazing angle geometry simplifies
the lensing geometry, effectively reducing it from a two dimensional problem to
one dimension.  The statistics of such reflections is sometimes called
glitter, and has many solvable
properties \citep{LonguetHiggins1960}.  This is illustrated in
Fig. \ref{fig:water_reflection}.


A similar effect occurs when the
observer is below the surface.  Two major distinctions arise: 1. the
waves can deform the surface to create caustics in projection. Near
caustics, Snell's law can lead to highly amplified refraction
angles\citep{2006ApJ...640L.159G}.  2. due to the odd image theorem, each caustic leads to two
images.  In an astrophysical context, the surface could be related to
magnetic reconnection 
sheets \citep{2015MNRAS.450.3201B}, which have finite widths to
regularize these singularities.  Diffusive structures have Gaussian
profiles, which were analysed in \citet{2012MNRAS.421L.132P}.  The
lensing details differ for convergent (under-dense) vs divergent
(over-dense) lenses, first considered by \citet{1998ApJ...496..253C}.

\begin{figure*}
\vspace{-0.8in}
\includegraphics[width=1.0\textwidth, angle=0]{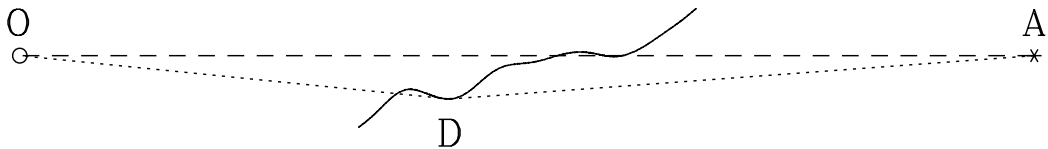}
\vspace{-7.0in}
\caption{Refractive lensing geometry 
(reproduced from \citet{2014MNRAS.442.3338P}
  fig. 1).  The pulsar is on the right, observer on the left.  Each
  fold of the sheet leads to a divergent projected density, resulting
  in a lensed image as indicated by the dotted line. See text for details.
}
\label{fig:sheetgeom}
\end{figure*}

The typical interstellar electron density $\sim 0.02$ cm$^{-3}$ 
is insufficient to deflect
radio waves by the observed $\sim$ mas bending angles.  At grazing
incidence, Snell's law results in an enhanced bending angle, which
formally diverges.  Magnetic discontinuities generically allow transverse surface waves, whose restoring force is the difference in Alfv\'en
speed on the two sides of the discontinuity.  This completes the
analogy to waves on a lake: for sufficiently inclined sheets the waves
will appear to fold back onto themselves in projection on the sky.  At
each fold caustic, Snell's law diverges, leading to enhanced
refractive lensing.  The divergence is cut off by the finite width of
the sheet.  The generic consequence is a series of collinear images.
Each projected fold of the wave results in two density caustics.  Each density
caustic leads to two geometric lensing images, for a total of 4 images
for each wave inflection.  The two geometric images in each caustic are
separated by the characteristic width of the sheet. If this is smaller
than the Fresnel scale, the two images become effectively
indistinguishable. 
The geometry of the inclined refractive lens is shown in Fig. \ref{fig:sheetgeom}.

A large number of sheets might intersect the line of sight to any
pulsar.  Only those sufficiently inclined would lead to caustic
formation.  Empirically, some pulsars show scattering that appears to be dominated by a
single sheet, leading to the prominent inverted
arclets in the secondary spectrum of the scintillations \citep{2001ApJ...549L..97S}.

\subsection{Archival data of B0834+06}
\label{21}
\begin{figure}
\centering
\includegraphics[width=\linewidth]{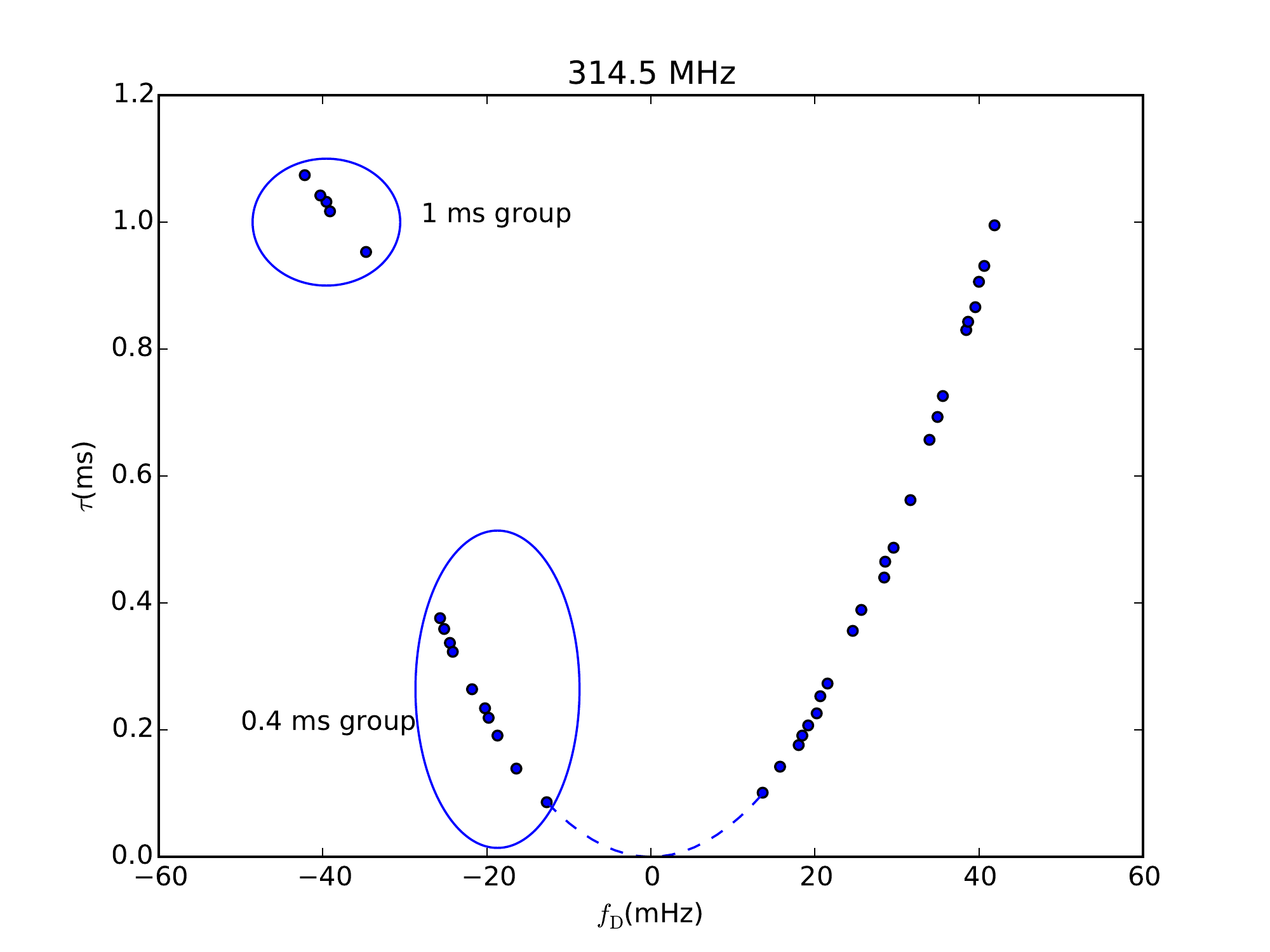}
\caption{Distribution of the apex positions in the sub-band centred at
  314.5 MHz.  The apexes that belong to the 1 and 0.4 ms groups are marked.}
\label{fig:apex_pos}
\end{figure}

Our analysis is based on the apex data selected from the secondary
spectrum of pulsar B0834+06 in B10, which was
observed as part of a 300 MHz global VLBI project on 2005 November 12, with
the GBT (GB), Arecibo (AR), Lovell and Westerbork (WB) telescopes.  The GB-AR and AR-WB
baselines are close to orthogonal and of comparable lengths, resulting
in relatively isotropic astrometric positions.
Information from each identified apex includes delay $\tau$,
delay rate (differential frequency $f_{\rm D}$), relative Right Ascension,
$\Delta\alpha$, and relative declination, $\Delta\delta$.
Data of each apex are collected from four dual circular polarization $8$ MHz wide sub-bands spanning the frequency range $310.5$--$342.5$ MHz. 
As described in B10, the inverse parabolic
arclets were fitted to positions of their apexes, resulting in a
catalogue of apexes in each sub-band, each with delay and differential
frequency.  
In this work, we first combine the
apexes across sub-bands, resulting in a single set of images.  We focus on
the southern group with negative differential frequency: this
grouping appears as a likely candidate for a double-lensing screen.  However, two groups (with negative differential frequency) appear distinct in both the VLBI angular positions and the secondary spectra.  We divide the apex data with negative $f_{\rm D}$ into two
groups: in one group, time delays range from $0.1$ ms to $0.4$ ms,
which we call the $0.4$ ms group; and in the other group, time delay at
about $1$ ms, which we call the $1$ ms group.  In summary, the
$0.4$ ms group contains $10$ apexes in the first two sub-bands, and
$14$ apexes in the last two sub-bands; the $1$ ms group, contains $5$,
$6$, $5$ and $4$ apexes in the four sub-bands subsequently, with
center frequency of each band $f_{\rm band}=314.5, 322.5, 330.5$ and $
338.5$ MHz.  The apex positions in the secondary spectrum are shown
in Fig. \ref{fig:apex_pos}. 

We select the equivalent apexes from four sub-bands.  To match the same apexes in different sub-bands, we scale the differential frequency in different sub-bands to $322.5$ MHz, by $f_{\rm D} (322.5/f_{\rm band})$ MHz.  A total of $9$ apexes from the $0.4$ ms group and $5$ apexes from the $1$ ms
group, were mapped.  This results in an estimation
for the mean referenced frequency $f=322.5$ MHz and a standard
deviation among the sub-bands, listed in Table
\ref{table:apex}.  The $f_{\rm D}$, $\tau$, $\Delta\alpha$ and $\Delta\delta$
are the mean values of $n$ sub-bands
($n=3$ for points 4 to 6 and points $1^\prime$, $2^\prime$ and $4^\prime$, while 4 for the remainder of the points), listed
in Table \ref{table:apex}. 

\begin{table*}
\centering
\begin{tabular}{c|llllll}
\hline
label & $\theta_{\parallel}$(mas) & $f_{\rm D}$(mHz) & $\tau$(ms)  & $\Delta\alpha$(mas) & $\Delta\delta$(mas) & $t_0$(day)\\
\hline
1& $-17.22$   & $-26.1(4)$    & 0.3743(6)         & 6.2    & $-11.9$      & $-107$                               \\
2& $-16.36$  & $-24.9(4)$      & 0.3378(3)         & 8.0(4)  & $-14.5(8)$      &$-101$                                \\
3& $-16.08$   & $-24.6(4)$       & 0.327(3)   & 7.2(6)  & $-13.9(4)$       & $-99.0$                                \\
4& $-14.45$   & $-22.3(5)$      & 0.2633(3)    & 6.1(4)  & $-13.1(7)$     & $-88.1$                                \\
5& $-13.68$ & $-21.6(6)$        & 0.236(2)    & 5.1(4)  & $-12.7(5)$      & $-83.3$                                \\
6& $-13.27$ & $-20.4(5)$      & 0.222(3)     & 5.8(4)  & $-11.8(1)$    & $-81.4$                                \\
7& $-12.21$   & $-18.9(2)$      & 0.188(2)   & 5.5(6) & $-10.8(6)$      & $-74.2$                        \\
8& $-10.58$   & $-16.8(3)$      & 0.1412(9)  & 3.9(6) & $-10.0(4)$      & $-62.8$                                \\
9& $-8.18$   & $-12.9(2)$      & 0.0845(5) & 2.8(3)  & $-8.6(4)$      & $-48.7$                                
\\ \hline

1'&$ \cdots$ & $-43.1(4)$      & 1.066(5)    & $-8(3)$     & $-24(2)$   & $-185$   \\
2'&$\cdots$ & $-41.3(5)$    & 1.037(3)   & $-14(1)$     & $-23(3)$   & $-188$                                   \\
3'&$\cdots$ & $-40.2(6)$  & 1.005(8)    & $-14(1)$       & $-22.3(5)$  & $-187$                                   \\
4'&$\cdots$ & $-38.3(6)$  & 0.9763(9)   & $-14(1)$       & $-20.6(3)$  & $-190$                                    \\
5'&$\cdots$& $-35.1(5)$  & 0.950(2)   & $-15(1)$  & $-21(1)$   &$-202$                                   \\
 \hline                                 
\end{tabular}
\caption{ $0.4$ ms and $1$ ms reduced apex data. 
We list the $0.4$ ms group data in the upper part of the table, while the $1$ ms group lie in the lower part of the table. 
Observation data include the differential frequency $f_{\rm D}$, time delay
$\tau$ ($\tau_1$ for $0.4$ ms group and $\tau_2$ for $1$ ms group);
$\Delta\alpha$ and $\Delta\delta$ are from the VLBI measurement (there is only one matched position for point 1, thus no error). 
$t_0$ is the time at constant velocity for an apex to intersect the
origin at constant speed along the main scattering parabola.  More
details in Section \ref{21} and Section \ref{222}.
}
\label{table:apex}
\end{table*}
We estimate the error of time delay $\tau$, differential frequency
$f_{\rm D}$, $\Delta\alpha$ and $\Delta\delta$  listed in Table
\ref{table:apex} from their band-to-band variance:
\begin{equation}
\sigma^2_{\rm \tau, f_{\rm D},\Delta\alpha, \Delta\delta} = \frac{1}{n}\sum^{n}_{i=1}\frac{(x_i-\bar{x})^2}{n-1},
\end{equation}
and $n$ is the number of sub-bands. The outer $1/n$ accounts for the expected variance of a mean of $n$ numbers.


\subsection{One-lens model}
\subsubsection{Distance to the lenses}
In the absence of a lens model, the
fringe rate, delay and angular position cannot be uniquely related.  To interpret the data, we adopt the lensing model of
\citet{2014MNRAS.442.3338P}.  In this model, the lensing is due to projected fold caustics of a thin sheet closely aligned to the line of sight.  We will list the parameters in this lens model in Table \ref{tab:parameters}.

\begin{table}
\caption{Parameters for double-lens model}
\begin{threeparttable}
\begin{tabular}{cl}
\hline
$D_{1\rm e}$  & 
Effective Distance of $0.4$ group data\\
$D_{2\rm e}$ 
	&  Effective Distance of $1$ ms group data\\
$D_1$ 		&  Distance of lens 1 \\
$D_2$	& Distance of lens 2 \\
$\gamma$ &  Scattering axis angle of $0.4$ ms group\tnote{a} \\
$\phi$	& Angle of the velocity of the pulsar\tnote{a}\\
$\theta$ & Angular offset of the object \\
\hline
\end{tabular}

\begin{tablenotes}
\item[a]{The angle is measured relative to the longitude and east is the positive direction.}
\end{tablenotes}
\label{tab:parameters}
\end{threeparttable}
\end{table}

\begin{figure}
\centering
\hspace*{-0.4in}\includegraphics[width=3.9in]{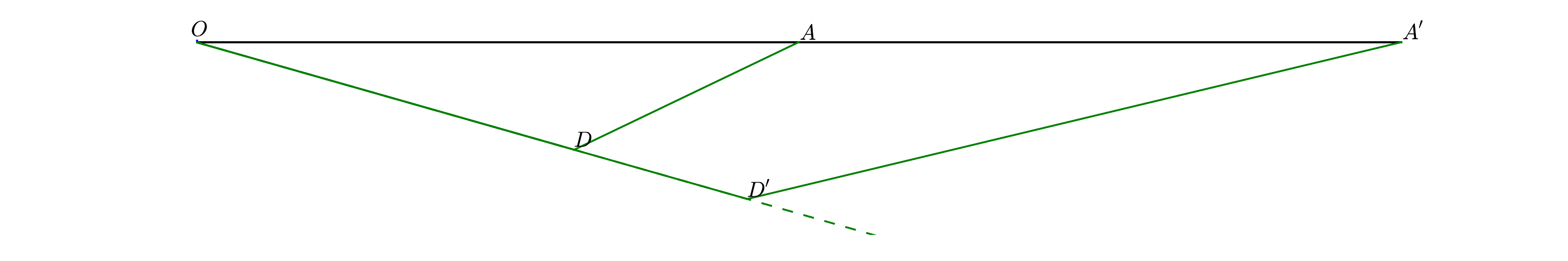}
\caption{A single refracted light path showing the distance
  degeneracy.  The primed and un-primed geometries result in the same
  observables: delay $\tau$ and angular offset $\theta$.
$O$ denotes the observer; $A$ and $A'$ denote the positions of the
pulsar; $D$ and $D'$ denote the positions of the refracted images on the
interstellar medium. The un-primed geometry corresponds to a pulsar
distance $D_{\rm p} = |AO| = 620$ pc, while the primed geometry has the same $D_{\rm e}$ but twice the
$D_{\rm p}.$}
\label{fig:Singledegeneracy}
\end{figure}

We define the {\it effective distance} $D_{\rm e}$ as
\begin{equation}
D_{\rm e} \equiv \frac{2c\tau}{\theta^2}.
\end{equation}
The differential frequency is related to the rate of change of delay
as $f_{\rm D}  =-f\frac{\rm d\tau}{\dif t}$.    
In general, $D_{\rm e}={D_{\rm p} D_{\rm s}}/({D_{\rm p} - D_{\rm
    s}})$ for a screen at $D_{\rm s}$.  
    The effective distance
corresponds to the pulsar distance $D_{\rm p}$, if the screen is exactly halfway.  
Fig. \ref{fig:Singledegeneracy} shows two sets of $D_{\rm p}$ and $D_{\rm s}$ with common $D_{\rm e}$.

\begin{figure}
\centering
\includegraphics[width=1.0\linewidth, angle=0]{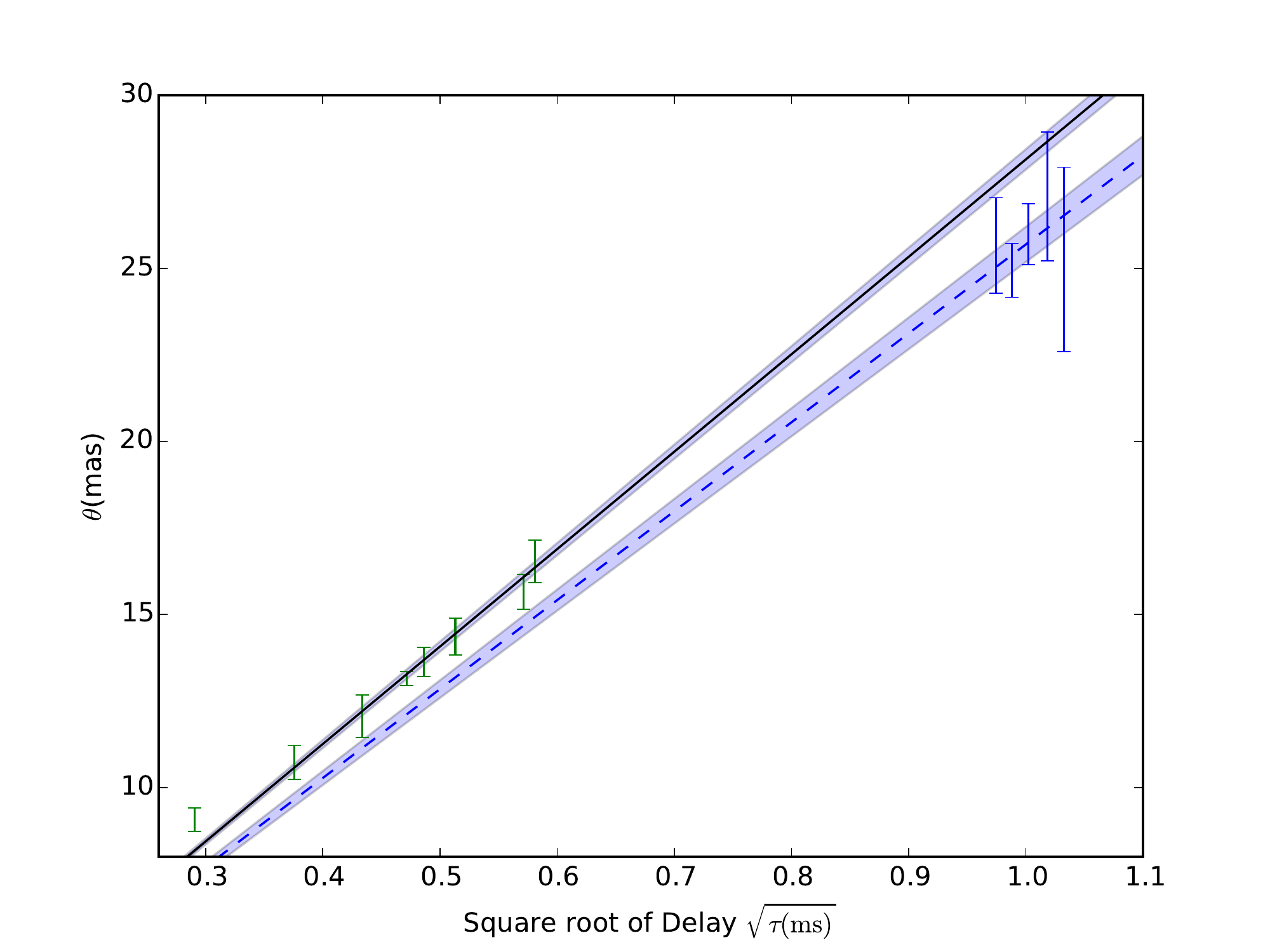}
\caption{${\theta}$ vs ${\sqrt{\tau}}$.  Two separate lines through the
  origin were fitted to the points sampled among the $0.4$ ms group
  and $1$ ms group.  The solid line is the fitted line of the $0.4$ ms
  positions, where $D_{1\rm e}=1044\pm 22$ pc. The dashed line is
  the fitted line of the $1$ ms 
  position, where $D_{2\rm e}=1252\pm 49$ pc.
}
\label{thetatau}
\end{figure}

When estimating the angular offset of each apex,
we subtract the expected noise bias:
${\theta}^2=({\Delta\alpha}\cos(\delta))^2+({\Delta\delta})^2-\sigma^2_{\Delta\alpha}-\sigma^2_{\Delta\delta}$. 
We plot the $\theta$ vs square root of $\tau$ in Fig.
\ref{thetatau}.  A least-square fit to the distance results in
$D_{1{\rm e}}=1044\pm 22$ pc for the  $0.4$\ ms group, which we call
lens 1 (point 1 is excluded since VLBI astrometry was only known for
one sub-band, thus we cannot obtain the variance nor weighted mean for that point), and
$D_{2{\rm e}} = 1252 \pm 49$ pc for the $1$\ ms group, hereafter lens 2.
The errors, and uncertainties on the error, preclude a definitive
interpretation of the apparent difference in distance.  However, at face value, 
this indicates that lens 2 is closer to the pulsar, and we will
use this as a basis for the model in this paper.  
The distances are slightly different from those derived in
B10, which is partly due to a different subset
of arclets analysed.  We discuss
consequences of alternate interpretations in Section \ref{sec:degeneracy}.
The pulsar distance was directly measured using VLBI parallax to be
$D_{\rm p} = 620 \pm 60$ pc, described in more detail in Section \ref{sec:astrometry}.  
We take $D_{1{\rm e}}=1044$ pc,  and the distance of lens 1 $D_{1}$,
where $0.4$ ms group scintillation points are refracted, as $389$
pc. Similarly, for $1$ ms apexes, the distance of lens 2 is taken as $D_2=415$ pc,
slightly closer to the pulsar.

\begin{figure*}
\centering
\includegraphics[width=7.5in]{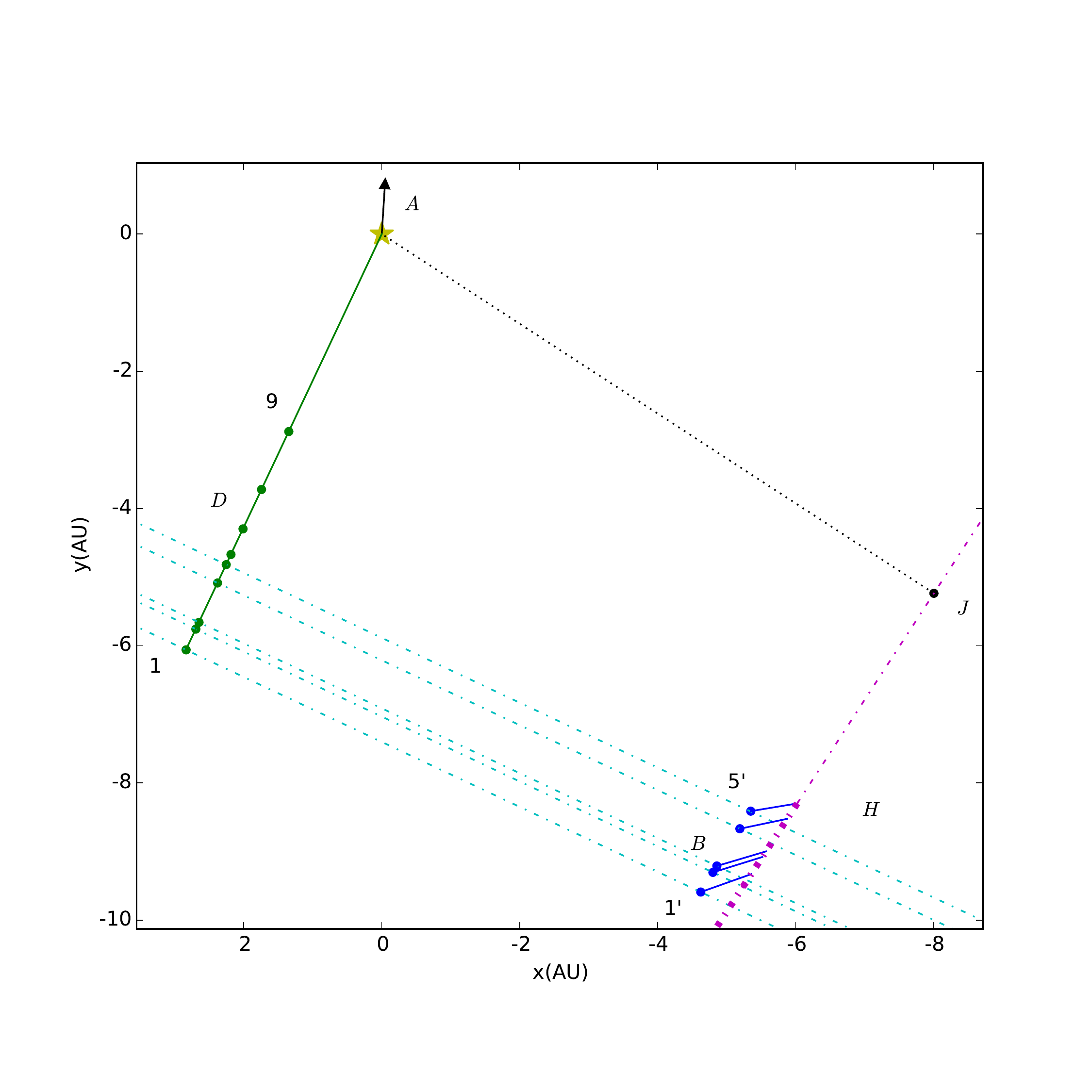}
\caption{Positions of $0.4$ ms and $1$ ms group data in the one-lens model and double-lens model.  
The axes represent the relative transverse distances to the un-refracted pulsar in Right Ascension (calculated by $x= \Delta \alpha \cos(\delta)D_i$ with $D_i$ represent the distance of the object to the observer: for point $A$, $D_i=D_{\rm p}$; for points $H$ and $J$, $D_i=D_2$; for points $B$ and $D$, $D_i=D_1$.) and declination (calculated by $y=\Delta \delta D_i$) directions, on a 2D plane that is transverse to the line of sight. 
On the left side, the points marked with letter $D$ labelled from 1 to 9, are the derived
positions from the time delays of $0.4$ ms group in 
the one-lens
model.  At a distance $389$ pc from the observer, the green solid line
demarcates the scattering axis for the
$0.4$ ms apexes positions, with an angle $\gamma=-25\degree.2$ east of
north.  The points on the right side mark the first and second
refraction points in double-lens model.  
The unobservable points
denoted by the letter $H$, are the calculated positions on lens 2 from the
1
 ms group; the observed apparent positions denoted by the letter $B$, are the  
second refraction on lens 1.  They are connected by short solid lines.  The long
dash dotted line passing through $J$ is the inferred geometry of the second
lens.  Its thicker portion has formed a full caustic, while the
thinner portion are sub critical.
The dash 
dotted lines, constructed perpendicular to the $AD$ scattering axis,
denote the caustics of lens 1.  The dotted line on the top right is
perpendicular to the magenta dash dotted line, intersecting at $J$.   
The relative model pulsar-screen velocity is $185.3$ km $\rm s^{-1}$, with an angle $\phi=-3\degree.7$ east of north, is marked with an arrow from the star, at point $A$, at the top of the figure.} 
\label{Doublelens}
\end{figure*}

For the 0.4 ms group, we adopt the geometry from
B10, assigning these points along line $AD$ as
shown in Fig. \ref{Doublelens} based solely on their delay, which is
the best measured observable.  The line $AD$ is taken as a
fixed angle of $\gamma=-25\degree.2$ east north.  We use this axis
to define ${\parallel}$ direction and define ${\bot}$ by a $90\degree$ direction clockwise
rotation.

\subsubsection{Discussion of one-lens model}
\label{222}
The $0.4$ ms group lens solution appears consistent with the premise
of the inclined sheet lensing model \citep{2014MNRAS.442.3338P}, which
predicts collinear positions of lensing images.  The time in the last
column of Table \ref{table:apex}, which we denote as $t_0
=-2{\tau}f/{f_{\rm D}}$, corresponds to the time required 
for the arclet to drift in the secondary spectrum through a delay of zero.

The collinearity can be considered a post-diction of this model.  The
precise positions of each image are random, and with 9 images no
precision test is possible.  The predictive power of the sheet model
becomes clear in the presence of a second, off-axis screen, which will be discussed below.



\subsection{Double-lens model}
\label{doublelensmodel}

The apparent offset of the $1$ ms group can be explained by a second lens
screen.  The small number of apexes at $1$ ms suggests that the second
lens screen involves a single caustic at a different distance.  One
expects each lens to re-image the full set of first scatterings,
resulting in a number of apparent images equal to the product of
number of lenses in each screen.
In the
primary lens system, the inclination appears such that typical waves
form caustics.  For the sake of discussion, we consider an inclination
angle for lens 1 $\iota_1 = 0.1^o$, and a typical slope of waves
$\sigma_\iota = \iota_1$.   Each wave of gradient larger than
1-$\sigma$ will form a caustic in projection. 
The number of sheets at shallower inclination
increases as the square of this small angle.  A 3 times less-inclined $\iota_2=0.3^o$
sheet occurs 9 times as often.  For the same amplitude waves on this
second surface, they only form
caustics for 3-$\sigma$ waves, which occur two hundred times less
often.  Thus, one expects such sheets to only form isolated caustics,
which we expect to see occasionally.  Three free parameters describe a
second caustic: distance, angle, and angular separation.  We fix the
distances from the effective VLBI distance ($D_1$ and $D_2$), and fit the angular
separations and angles with the 5 delays of the $1$ ms group.

\subsubsection{Solving the double-lens model}

Apexes 1'--5' share a similar 1 ms time delay, suggesting they are lensed by
a common structure.
We denote the position of the pulsar point as point $A$, the positions of the lensed image on lens 2 as point $H$, positions of the lensed image on lens 1 as point $B$, position of the observer as point $O$, and the nearest point on lens 2 to the pulsar as point $J$.  
The lines $AJ\perp HJ$ intersect at point $J$, $HF\perp BD$ intersect at the point $F$, and $BG\perp HJ$ intersect at the point $G$.

\begin{figure*}
\hspace*{-1in}\includegraphics[width=9in]{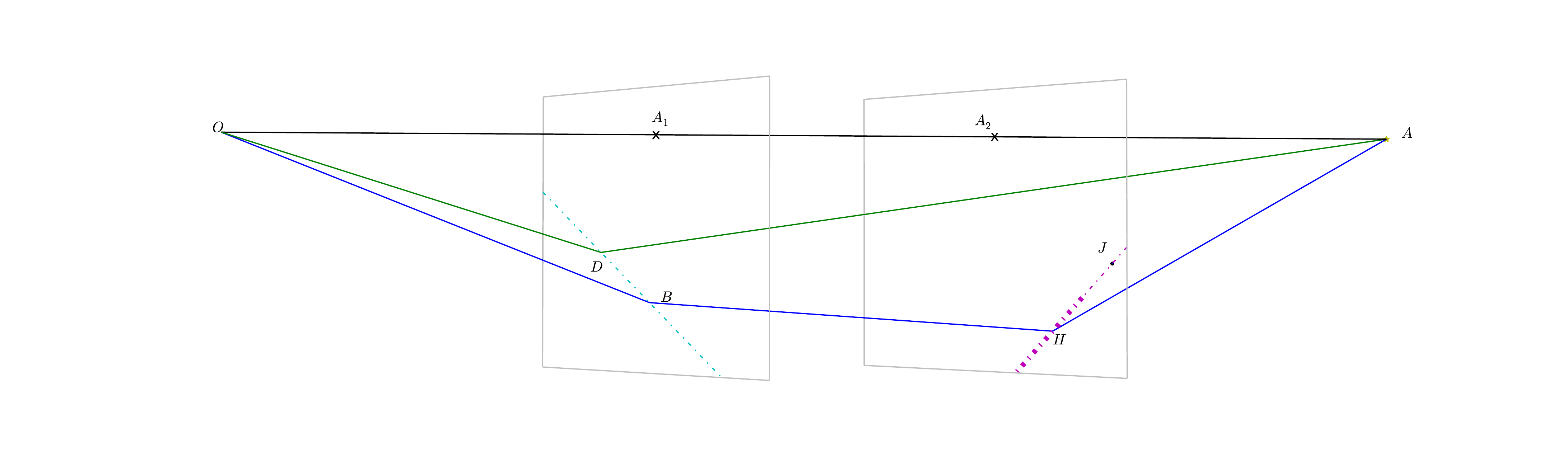}
\caption{A 3D schematic of light path when light is doubly-refracted.  
Two planes from left to right are the plane of lens 1,
  and the plane of lens 2.  The dash dotted lines represent the
  projected line-like fold caustics. 
Light goes from $A$ (pulsar) to the first
  refracted point $H$ on lens 2 (line $HJ$, magenta dash dotted line), and
  then the second refracted point $D$, the image we observe, on lens 1
  (line $BD$, cyan dash dotted line), and finally the observer $O$.  Dash
  dotted lines show the light path of singly-deflected light path
  ($A-D-O$).  The crosses ($A_1$ on plane 1, and $A_2$ on 
  plane 2) denote intersection of the un-deflected light through the
  lensing sheet.  $D$ and $J$ are the closest point of the lens caustic
  to the un-deflected path, which are the loci of single deflection
  images.  Thus, $A_2J\bot
  HJ$, and $A_1D\bot BD$.  The thick line on plane 2 indicates the
  real caustic, while the thin continuation indicates the extrapolated
  continuation beyond the cusp/swallowtail.
}
\label{fig:3D_image}
\end{figure*}

A 3D schematic of two plane lensing by linear caustics is shown in
Fig. \ref{fig:3D_image}. 

First, we calculate the position of $J$.  We estimate the
distance of $J$ from the $1$ ms $\theta$--$\sqrt{\tau}$ relation (see
Fig. \ref{thetatau}).  We determine the position of $J$ by
matching the time delays of point 4' and point 1', which is
marked in Fig. \ref{Doublelens}.  The long dash dotted line on the right side of Fig. \ref{Doublelens}
denotes the inferred geometry of lens 2, and by construction vertical
to $AJ$. 

The second step is to find the matched pairs of those two lenses.  
By inspection, we found that the 5 furthest points in $0.4$ ms group match naturally to the double-lens images.  These five matched lines are marked with cyan dash dotted lines in Fig. \ref{Doublelens} and their values are listed in the second column in Table \ref{table:double_lens_compare}. 
\begin{table*}
\centering
\begin{tabular}{c|llllllll}
\hline
label&$\theta_{\parallel}$ (mas)  & $\tau_2$(ms) & $\sigma_{\rm \tau}$(ms)  & $\tau_{\rm M}$(ms) & $f_{\rm D}$(mHz)  &$\sigma_{f}$(mHz)      &  $f_{\rm M}$(mHz)& $t_1$(day) \\ \hline
1'& $-17.22$  &  1.0663     &0.0050    & 1.0663*        & $-43.08$    &0.84   & $-42.26$           & $-78$\\
 2'& $-16.36$  &    1.0370     &0.0059    & 1.0362       & $-41.27$    & 0.88   & $-41.04$          & $-73$\\ 
3'& $-16.08$  &   1.005    &0.011   & 1.027          & $-40.17$    &   0.87     & $-40.64$          & $-72$\\ 
 4'& $-14.45$  &   0.9763    &0.00088   & 0.9763*       & $-38.31$     &0.64    & $-38.31\dagger$  & $-63$\\ 
5'& $-13.68$  &    0.9495     &0.0094    & 0.9550       & $-35.06$     &0.78    & $-37.21$          &$-59$\\ 
 \hline
\end{tabular}
\caption{Comparison of time delay $\tau$ and the differential
  frequency $f_{\rm D}$ of the observation and the model fitting result in the
  double-lens model.  $\theta_{\parallel}$ denotes the angular offsets
  of the corresponding images at lens 1. 
The values with star symbols on them are the points that we use to
calculate the position of $J$ and the point with a $\dagger$ symbol is
the point that we use to calculate the transverse velocity of the
pulsar $v_{\bot}$.  They agree with data by construction.  The last
column, $t_1$ is the time the lensed image on lens 2 takes to move
from point $H$ to point $J$, which is also defined in Section
\ref{subsec:doublelens}.} 
\label{table:double_lens_compare}
\end{table*}
They are the located at a distance $389$ pc away from us. Here we define three distances:
\begin{equation}
\begin{aligned}
D_{\rm p2}&=620~{\rm pc}-415~{\rm pc} =205~{\rm pc},\\
D_{21}&=415~{\rm pc}-389~{\rm pc} =26~{\rm pc},
\end{aligned} 
\end{equation}
where $D_{\rm p2}$ is the distance from the pulsar to lens 2, and $D_{21}$ is the distance from lens 2 to lens 1.

\begin{figure}
\centering
\includegraphics[width=1.0\linewidth,scale=1.0]{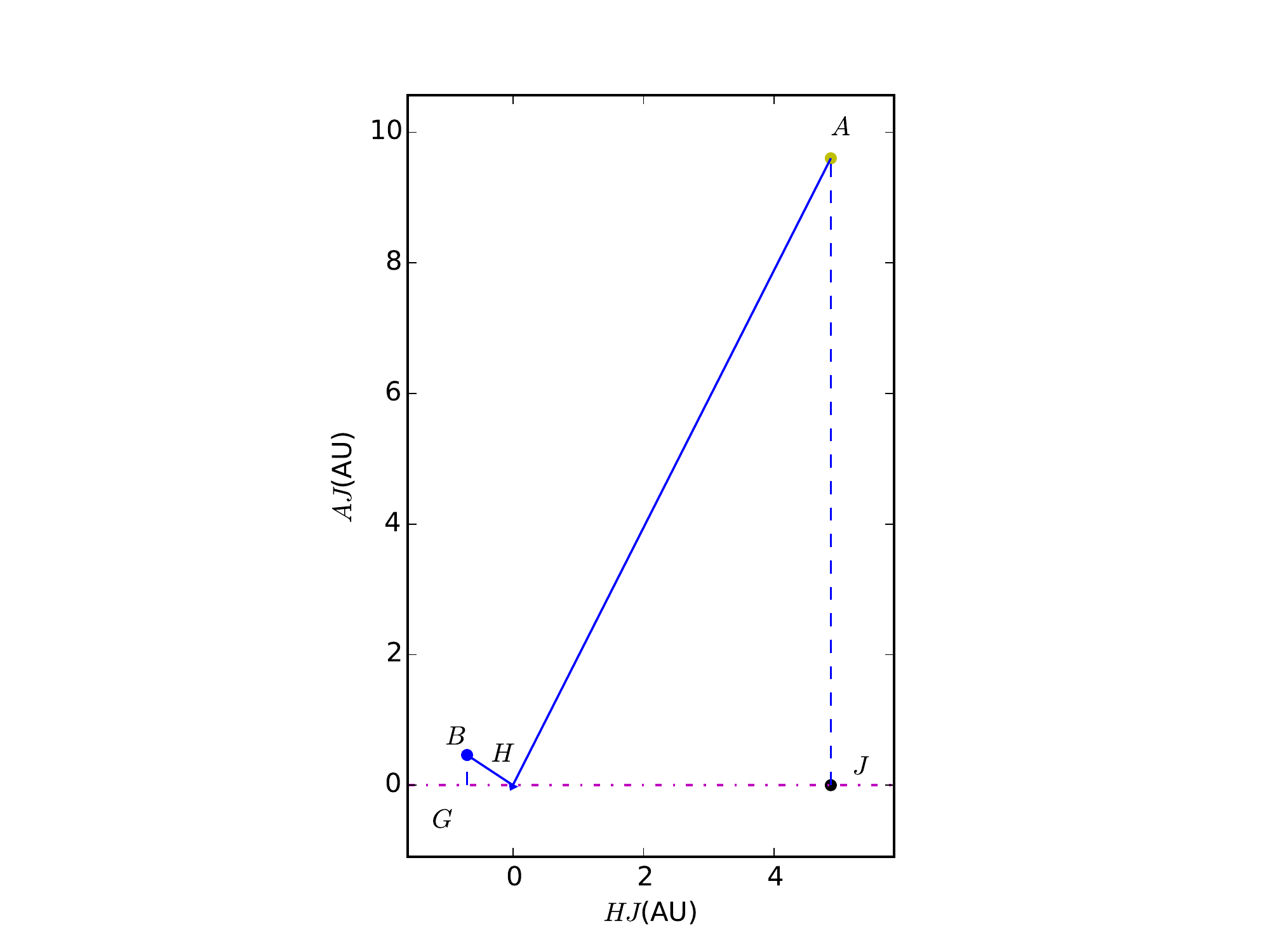}
\caption{Refraction on lens 2. 
$A$ is the position of the pulsar.  $H$ is the lensed image on lens 2.  $B$ is
the lensed image on lens 1.  $AJ\bot HJ$ and $BG\bot HJ$.  We illustrate the scenario for point 1'.}
\label{first_reflect}
\end{figure}

\begin{figure}
\centering
\includegraphics[width=1.0\linewidth]{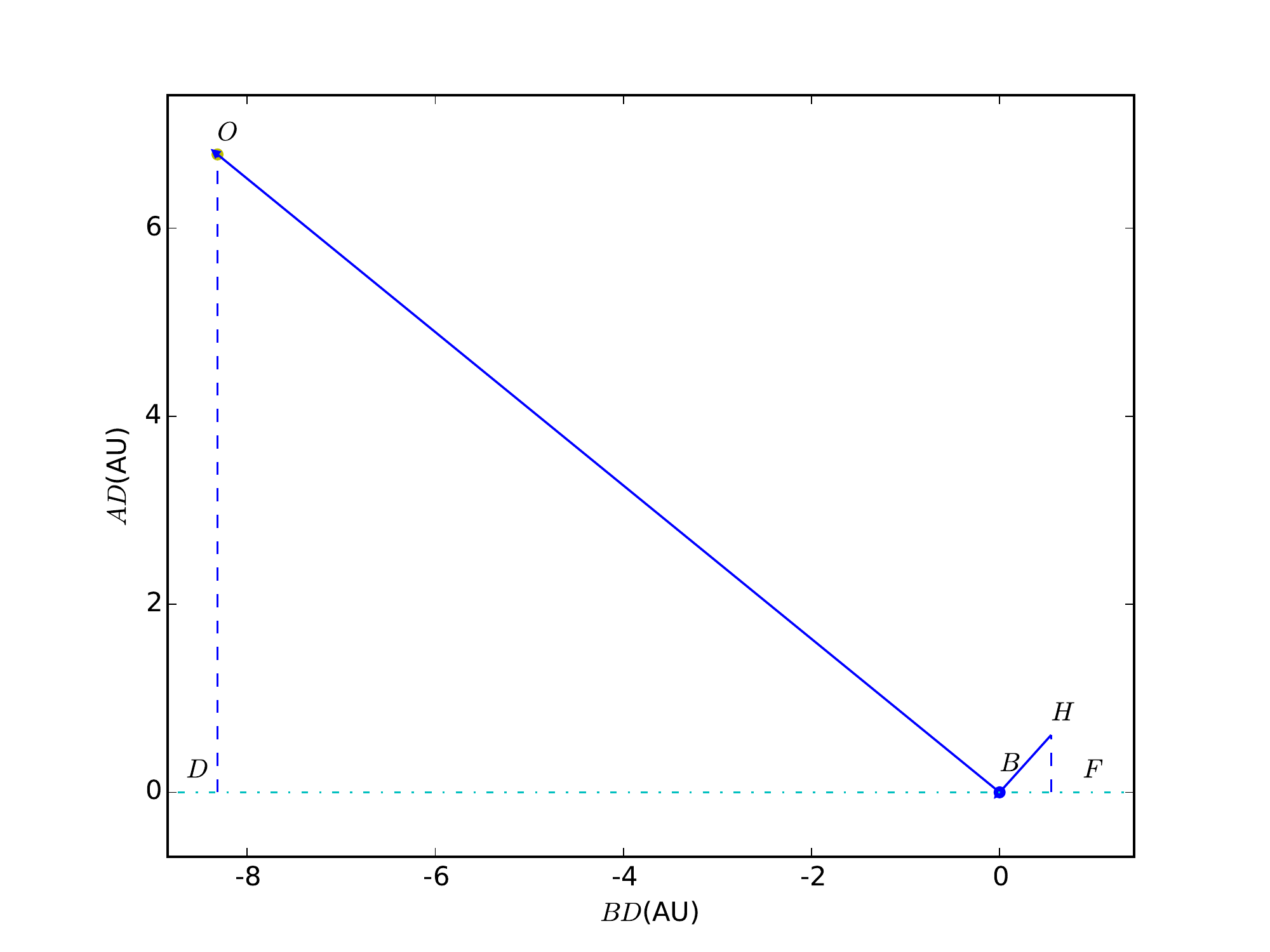}
\caption{Refraction on lens 1. 
$H$ is the lensed image on lens 2.  $B$ is the lensed image on lens 1.  $O$ is the position of the observer. $HF\bot DF$ and $OD\bot HJ$.  As in the previous figure, we illustrate the scenario for point 1'. }
\label{second_reflect}
\end{figure}

Fig. \ref{first_reflect} and Fig. \ref{second_reflect} are examples of how light is refracted on the first lens plane and the second lens plane.  We specifically choose the point with $\theta_{\parallel}=-17.22$ mas, which refer point 1' on lens 1 as an example.  
Equality of the velocity of the photon parallel to the lens plane before and after refraction implies the relation:
\begin{equation}
\begin{aligned}
\frac{JH}{D_{\rm p2}}&=\frac{HG}{D_{21}},\\
\frac{FB}{D_{21}}&=\frac{BD}{D_{1}}.
\end{aligned}
\end{equation}

We plot the solved positions in Fig. \ref{Doublelens}, and list
respective time delays and differential frequencies in
Table \ref{table:double_lens_compare}.  
We take the error of the time delay $\tau$
in the double-lens model as
\begin{equation}
\begin{aligned}
(\frac{\sigma_{\tau_i}}{\tau_{2i}})^2 = (\frac{\sigma_{\tau1i}}{\tau_{1i}})^2+(\frac{\sigma_{\tau2i}}{\tau_{2i}})^2 + (\frac{\sigma_{\tau2j}}{\tau_{2j}})^2,
\end{aligned}
\end{equation}
where $\tau_1$ and $\sigma_{\tau1}$ represent the time delay and its
error from the $0.4$ ms group on lens $1$, and $\tau_2$ and
$\sigma_{\tau2}$ represent the time delay and its error from
the $1$ ms group on lens 2.  And $\tau_{2j}$ is the $\tau_2$ for
the nearest reference point in Table
\ref{table:double_lens_compare} with error $\sigma_{\tau2j}$.  
Specifically,
for point $i=5'$ and $3'$, $j=4'$ is the nearest reference point; while for point $i=2'$, $j=1'$ is the nearest reference point.  The reference points are marked with star symbols in the fifth column in Table \ref{table:double_lens_compare}.

For the error of differential frequency $f_{\rm D}$, we add the error of the reference point (point 4') to the error of each other measured point:
\begin{equation}
\begin{aligned}
(\frac{\sigma_{f_i}}{f_{{\rm D}i}})^2=(\frac{\sigma_{f_{{\rm D}i}}}{f_{{\rm D}i}})^2+(\frac{\sigma_{f_{{\rm D}4'}}}{f_{{\rm D}4'}})^2
\end{aligned}
\end{equation}
where $f_{{\rm D}4'}$ and $\sigma_{f_{{\rm D}4'}}$ are the differential frequency
and we list its error of the point in the fourth row in Table
\ref{table:double_lens_compare}.

\subsubsection{Comparing with observations}
In order to compare $\tau$, we calculate model time delays $\tau_{\rm M}$ for these five
points, and list the results in Table
\ref{table:double_lens_compare}.  For points 4' and 1', they fit by
construction since we use these to calculate the position of $J$; for the remaining three points, all of the results are within 3-$\sigma$ of the observed time delays.

To compare differential frequency $f_{\rm D}$, we need to calculate the velocity of the pulsar and the velocity of the lens. We take the lenses to be static, and solve the velocity of the pulsar relative to the lens (in geocentric coordinates).  The pulsar has two velocity components, and the two 1-D lenses effectively determine one component each.
For $v_{\parallel}$, we derive the velocity $172.4 \pm 2.4$ km $\rm s^{-1}$, 
which is $58.7$ mas/yr in a geocentric system, 
from $f_{\rm D}$ of point $1$ in $0.4$ ms group.  The direct observable is the time to crossing of each caustic, denoted $t_0$ in Table \ref{table:apex}. 

To calculate $v_{\bot}$, we choose the point 4', which has the smallest errorbar of differential frequency.
This gives a value of  $67.9\pm 2.8$ km $\rm s^{-1}$ for $v_{\perp}$,  which is $21.4$ mas/yr in geocentric system, with an angle
$\phi=-3\degree.7 \pm 0\degree.8$ west of north.  This represents the
pulsar-screen velocity relative to the Earth.  We can further
transform this into the local standard of rest (LSR) frame to
interpret the velocities in a Galactic context.
The model derived and observed velocities (heliocentric and LSR) are listed in Table \ref{Table:velocity}.  The
direction of the model velocity is marked on the top of the star in
Fig. \ref{Doublelens}. 


\begin{table*}
\centering
\begin{tabular}{c|cccccc}
\hline
Parameter & $\mu_{\alpha*}$(mas yr$^{-1}$) & $\mu_{\delta}$(mas yr$^{-1}$) & $\mu_{l*}$(mas yr$^{-1}$) & $\mu_b$ (mas yr$^{-1}$) & $v_{l*}$ (km $\rm s^{-1}$) & $v_b$ (km $\rm s^{-1}$) \\
\hline
model pulsar-screen velocity & $-5.30 \pm 1.11$  & $61.97 \pm 1.11$  & $-56.45$ & 22.23 &  $\cdots$ & $\cdots$\\
VLBI pulsar proper motion & $2.16 \pm 0.19$ & $51.64 \pm 0.13$ & $-46.69$ & 28.02 & -137.24 & 82.34 \\
Screen motion & $\cdots$ & $\cdots$  & 9.76 & 5.79 & 18.00 & 10.68\\
\hline
\end{tabular}
\caption{Summary of velocities in double-lens model.  The velocities
listed in equatorial coordinates are the relative velocity in
heliocentric system, while the velocities in Galactic coordinates are
the relative velocities in LSR (Local Standard of Rest). $\mu_{\alpha*}=\Delta\alpha\cos(\delta)/t$ and $\mu_{l*}=\Delta l\cos(b)/t$, for we moved the center position from $(\Delta\alpha,\Delta\delta)$ or $(l,b)$ to (0,0).  $v_{l*}$ and $v_b$ are the linear velocities relative to the LSR.  The screen is
only moving slowly ($\sim 21$ km $\rm s^{-1}$).  Ellipses reflect the
unobserved and frame dependent parameters.
}
\label{Table:velocity}
\end{table*}


With this velocity of the pulsar, we calculate the model differential frequency $f_{\rm M}$ of points 5',3',2' and 1'. Results are listed in Table \ref{table:double_lens_compare}. The calculated results all lie within the 3-$\sigma$ error intervals of the observed data. 

The reduced ${\chi}^2$ for time delay $\tau$ is $1.5$
for $3$ degrees of freedom
and $2.2$ for $f_{\rm D}$ for $4$ degrees of freedom.  This is consistent with the model.


Within this lensing model, we can test if the 
caustics are parallel.  Using the lag error range of double-lensed point 4 (the best
constrained), we find a 1-$\sigma$ allowed angle of 0.4 degrees from
parallel with the whole lensing system.  This lends support to the hypothesis of a 
highly inclined sheet, probably aligned to better than 1 per cent.


\subsubsection{Discussion of double-lens model}
\label{subsec:doublelens}
\begin{figure}
\centering
\includegraphics[width=1.0\linewidth]{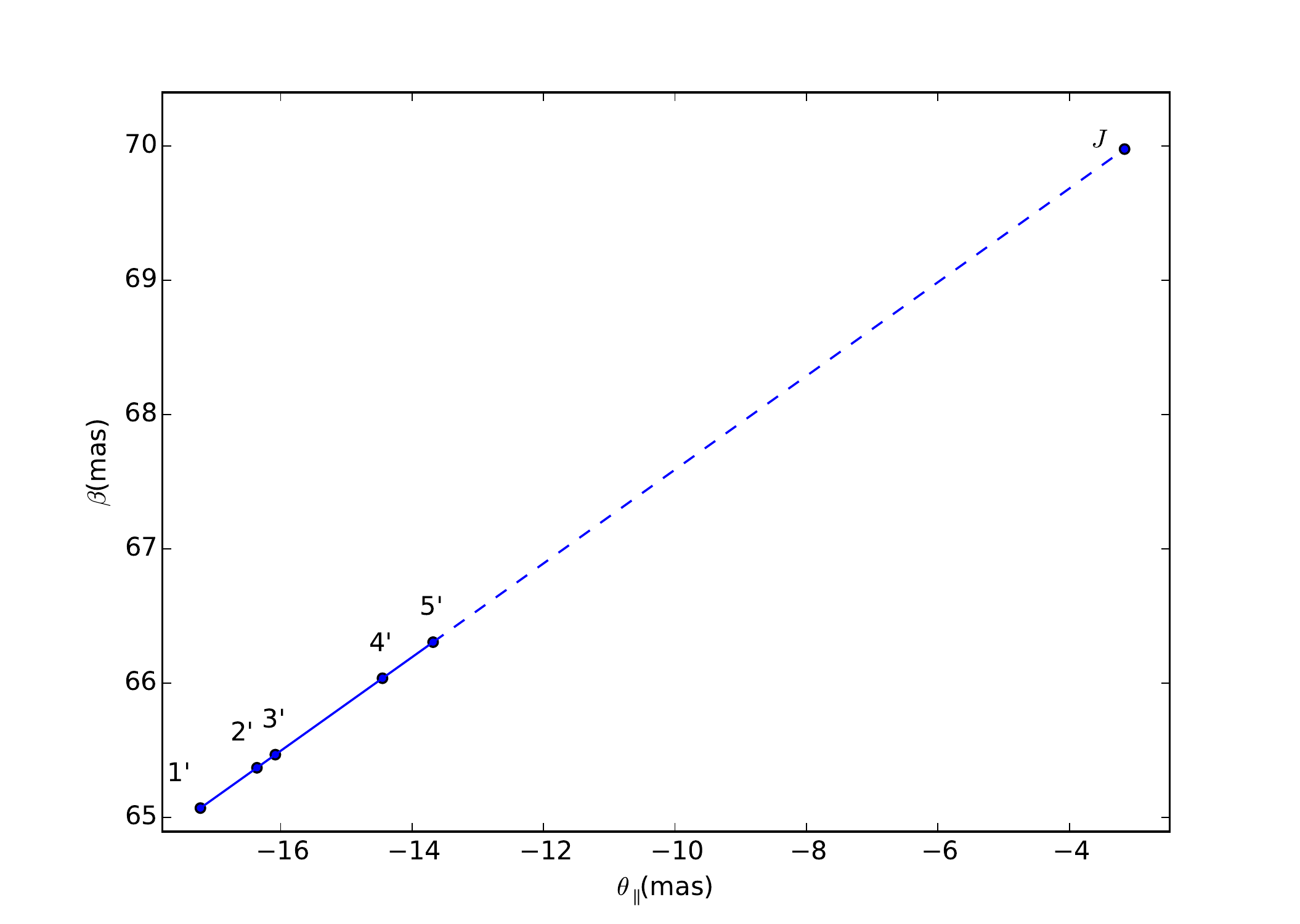}
\caption{Deflection angle $\beta=\pi-\protect\angle AHB$ on lens 2.  Point $J$ denotes the expected position to form a single refraction image, which was not
  observed.   The small change in angle relative to the observed
  images precludes a finite refraction cut-off, since the data spans
  10 per cent bandwidth, with a 20 per cent change in refractive strength.  We
  propose a swallowtail caustic as the likely origin for the
  termination of the second lens sheet.
}
\label{vtrans}
\end{figure}

For the $1$ ms group, lens 2
only images a subset of the lens 1 images.  This could happen if
lens 1 screen is just under the critical inclination
angle, such that only 3-$\sigma$ waves lead to a fold caustic.  If the lens 2 was at a critical angle, the chance of encountering a
somewhat less inclined system is of order unity.
More surprising is the absence of a single-refracted
image of the pulsar, which is expected at position $J$.  This could
happen if the maximum refraction angle is just below critical, such
that only rays on the appropriately aligned double refraction can form
images.  
We plot the refraction angle $\beta$ in the
direction that is transverse to the first lens plane in Fig.
\ref{vtrans}.  
The fractional bandwidth of the data is about 10 per cent, making it unlikely that single lens
image $J$ would not be seen due to the larger required refraction
angle.  Instead, we speculate that the fold caustic terminates
near double-lensed image 5', and thus only intersections with the closer
lens plane caustic south of image 5' are doubly-lensed.

This is a generic outcome of a swallowtail
catastrophe \citep{Arnold1990}.  In this picture, the sheet just
starts folding near point 5'.  North of point 5', no fold appears in
projection.  Far south of point 5', a full fold exhibits two caustics
emanating from the fold cusp.  Near the cusp the magnification is the
superposition of two caustics, leading to enhanced lensing and higher
likelihood of being observed.  

We denote $t_1$ the time for the lensed image on lens 2 to move from point
$H$ to point $J$.  From our calculation, we predict that on 2005 September 14, which is 59 days before the observation, the lensed image would have appeared overlayed on point $H_5$; and on 2005 August 26, which is 78 days before the observation, the lensed image would have appeared overlayed on point $H_1$.
The
model predicts the presence of a singly-lensed image refracted at
these points, in addition to the doubly-lensed images.

The generic flux of a lensed image is the ratio of the lens transverse
size to maximum impact parameter \citep{2012MNRAS.421L.132P}.  Near
the caustic, the lensed flux can become very high.  The 1 ms group is
about a factor of 4 fainter than the 0.4 ms group.  The high flux of
the second caustic suggests it to be relatively wide, perhaps a
fraction of an AU.  Due to the odd image theorem, one generically
expects two distinct set of double lensed arcs.  We only see one
(generically the outer one), which places an upper bound on the
brightness of the inner image.  In a divergent
lens\citep{1998ApJ...496..253C}, the inner image is generically much
fainter, so perhaps not surprising. For a convergent Gaussian lens,
the two images are of similar brightness, but a more cuspy profile
will also result in a faint inner image.  In gravitational lensing,
the odd image theorem is rarely seem to hold, generally thought to be
due to one lens being very faint.

\begin{figure}
\centering
\includegraphics[width=1.0\linewidth, angle=0]{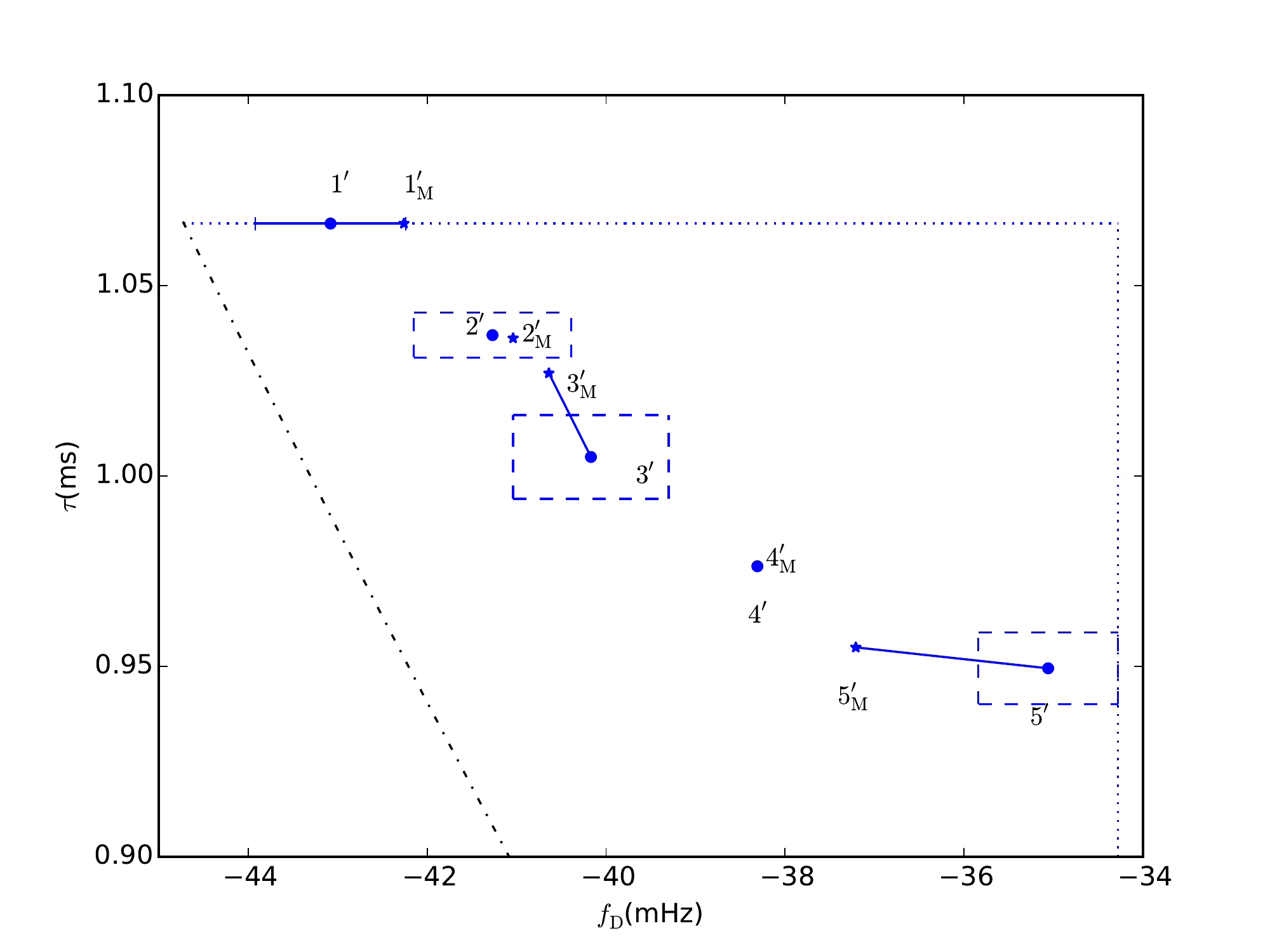}
\caption{Model comparison.
Points with subscripts are derived from the double-lens model, see
Table \ref{table:double_lens_compare}.  
Rectangles mark the 1-$\sigma$ error region.
Points 1' (only for $\tau$) and 4' were used to fit the model, and
thus do not have an error region.  The rectangles cover $10^{-3}$ of
the area in the dotted region bounded by the parabolic arc and the
data points.  We interpret this precise agreement
between model and data 
is unlikely to be a random coincidence.
}
\label{fig:tau-fD}
\end{figure}

One can try to estimate the chance of accidental agreement between
model and data.
We show the data visually in Fig. \ref{fig:tau-fD}.  

To estimate where points might lie accidentally, we conservatively
compare the area of the error regions to the area bounded by the
parabola and the data points, as shown by dotted lines.  This results
in about 10$^{-3}$, suggesting that the model is unlikely to be an
accidental fit.

\subsection{Distance degeneracies}
\label{sec:degeneracy}
\begin{figure*}
\centering
\hspace*{-1in}\includegraphics[width=9in]{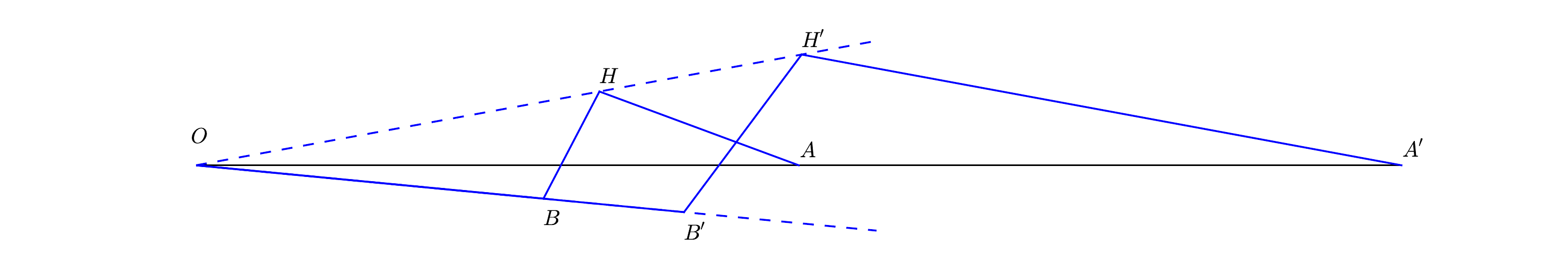}
\caption{Illustration of double-lens degeneracy.  As in
  Fig. \protect{\ref{fig:Singledegeneracy}}, all observables are identical
  for both the prime and un-primed geometries, including all pairwise
  delays and angular positions.  This degeneracy also holds in three dimensions.}
\label{fig:double_degeneracy}
\end{figure*}

With two lens screens, the number of observables increases: in
principle one could observe both single refraction delays and angular
positions, as well as the double reflection delays and angular
positions.  Three distances are unknown, equal to the number of
observables.  Unfortunately, these measurements are degenerate, which
can be seen as follows.  From the two screens $i=1,2$, the two single
deflection effective distance observables are
$D_{i{\rm e}} \equiv 2c\tau_i/\theta_i^2=D_i^2(1/D_i+1/D_{{\rm p}i})$.  A third
observable effective distance is that of screen 2 using screen 1 as a
lens, $D_{21{\rm e}}=D_1^2(1/D_1+1/D_{21})$, within the triangle that is formed by lens 1, lens 2 and the observer.  That is also algebraically
derivable from the first two relations:
$D_{{21}{\rm e}}=D_{1{\rm e}}D_{2{\rm e}}/(D_{2{\rm e}}-D_{1{\rm e}})$.  
We illustrate the light path in Fig. \ref{fig:double_degeneracy}.


In this archival data set, the direct single lens from the further
plane at position $J$ is missing.  It would have been visible $59$ days
earlier.  The difference in time delays to image $J$ and the double
reflection images would allow a direct determination of the effective
distance to lens plane 2.  Due to the close to $90 \degree$ angle $\angle~DAJ$
between lenses, the effect would be about a factor of 10 ill
conditioned.  With sufficiently precise VLBI imaging one could
distinguish if the doubly-refracted images are at position $B$ (if
lens 1 is closer to the observer) or position $H$ (if lens 2 is closer to the observer).  As described above, we interpret
the effective distances to place screen 2 further away.

\section{VLBI astrometry}
\label{sec:astrometry}

The model kinematics can be compared to direct measurements of pulsar
proper motion to infer the absolute motion of the lensing screen.

PSR B0834+06 was observed 8 times with the Very Long Baseline Array (VLBA), under the project code BB269, between 2009 May and 2011 January.  Four 16~MHz bands spread across the frequency range 1406 -- 1692 MHz were sampled with 2 bit quantization in both circular polarizations, giving a total data rate of 512 Mbps per antenna.  The primary phase calibrator was J0831+0429, which is separated from the target by 2.1 degrees, but the target field also included an in-beam calibrator source J083646.4+061108, which is separated from PSR B0834+06 by only 5 \arcmin.  The cycle time between primary phase calibrator and target field was 5 minutes, and the total duration of each observation was 4 hours.

Standard astrometric data reduction techniques were applied
\citep[e.g.,][]{deller12b,deller13a}, using a phase calibration
solution interval of 4 minutes for the in-beam calibrator source
J083646.4+061108.  J083646.4+061108 is weak (flux density $\sim$4 mJy)
and its brightness varied on the level of tens of percent.  The
faintness leads to noisy solutions, and the variability indicates that
source structure evolution (which would translate to offsets in the
fitted target position) could be present.  Together, these two effects
lead to reduced astrometric precision compared to that usually
obtained with VLBI astrometry using in-beam calibration, and the
results presented here could be improved upon if the observations were
repeated using the wider bandwidths and higher sensitivity now
available with the VLBA, potentially in conjunction with additional
in-beam background sources.

While a straightforward fit to the astrometric observables yields a pulsar distance with a formal error $<$1 per cent, the reduced $\chi^2$ of this fit is $\sim$40, indicating that the formal position errors greatly underestimate the true position errors, and that systematic effects such as the calibrator effects discussed above as well as residual ionospheric errors dominate.  Accordingly, the astrometric parameters and their errors were instead obtained by bootstrap sampling \citep{efron91a}.  These results are presented in Table~\ref{tab:vlbi}.

\begin{table}
\caption{Fitted and derived astrometric parameters for PSR B0834+06.}
\begin{threeparttable}
\begin{tabular}{ll}
\hline
Reference right ascension (J2000)\tnote{a}
							&  08:37:5.644606(9)\\
Reference declination (J2000)\tnote{a}
							&  06:10:15.4047(1)\\
Position epoch (MJD)			& 55200 \\
$\mu_{\mathrm{R.A.}}$ (mas yr$^{-1}$)	& 2.16(19) \\
$\mu_{\mathrm{Dec}}$	(mas yr$^{-1}$) & 51.64(13)  \\
Parallax (mas)	 				& 1.63(15) \\
Distance (pc)					& 620(60) \\
$v_{\mathrm T}$ (km s$^{-1}$)	& 150(15)\\
\hline
\end{tabular}
\begin{tablenotes}
\item[a]{The errors quoted here are from the astrometric fit only and do not include the $\sim$1 mas position uncertainty transferred from the in--beam calibrator's absolute position.}
\end{tablenotes}
\end{threeparttable}
\label{tab:vlbi}
\end{table}

\section{Discussions}
\label{sec:discussions}
\subsection{Interpretation}

The relative motion between pulsar and lens is directly measured by
the differential frequency, and not sensitive to details of this
model. B10 derived similar motions.  This
motion is in broad agreement with direct VLBI proper motion
measurement, requiring the lens to be moving slowly compared to the
pulsar proper motion or the LSR.  The lens is $\sim200$ pc above the
Galactic disk.  Matter can either be in pressure equilibrium, or in
free-fall, or some combination thereof.  In free fall, one expects
substantial motions.  These data rule out retrograde or radially
Galactic orbits: the lens is co-rotating with the Galaxy.  In pressure
equilibrium, gas rotates slower as its pressure scale height
increases, which appears consistent with the observed slightly slower
than co-rotating motion.  The modest lens velocities also appear
consistent with the general motion of the ISM, perhaps driven by
Galactic fountains \citep{1976ApJ...205..762S} at these latitudes above
the disk.  In the inclined sheet model, the waves move at Alfv\'enic
speed, but due to the high inclination, will move less than one percent of this
speed in projection on the sky, and thus be completely negligible compared to
other sources of motion.

Alternative models, for example, evaporating
clouds \citep{1998ApJ...498L.125W} or strange
matter \citep{2013PhLB..727..357P}, do not make clear predictions.  One
would expect higher proper motions from these freely orbiting sources,
and larger future scintillation samples may constrain these models.

In order to incline one sheet randomly to better than 1 per cent requires of
order $10^4$ randomly placed sheets, i.e. many per parsec.  This sheet
extends for $\sim$ 10 AU in projection, corresponding to a physical
scale greater than $1000$ AU.   These two numbers roughly agree,
leading to a physical picture of magnetic domain boundaries every
$\sim 0.1$ pc.  B0834+06 has had noted arcs for multiple years,
perhaps suggesting this dominant lens plane is larger than typical.
One might expect to reach the end of the sheet within decades.

A generic prediction of the inclined sheets model is a change in rotation
measure across the scattering length.  Over 1000 AU, one might expect
a typical RM (rotation measure) change of $10^{-3}$ rad/m$^2$.  At low frequencies, for
example in LOFAR\footnote{http://www.lofar.org/} or GMRT\footnote{http://gmrt.ncra.tifr.res.in/}, the size of the scattering screen extends
another order of magnitude in angular size, and the RM in different
lensed images are different, increasing to 
$\sim 0.01$, which is plausibly measurable.  Even for an un-polarized
source, the left and right circularly polarized (LCP, RCP) dynamic spectra will
be slightly different.  Usually a secondary spectrum (SS) is formed by
Fourier transforming the dynamic spectrum and multiplying by its
complex conjugate.  To measure the RM, one multiplies the Fourier
transform of the LCP dynamic spectrum by the complex conjugate of the
RCP Fourier transform.  This will display a phase gradient along the Doppler
frequency axis.  In the SS, each pixel is the sum of correlations of
pairs of scattering points with corresponding lag and Doppler
velocity.  The velocity is typically linear in the pair separation,
which is also the case for differential RM.   This statistic is
analogous to the cross gate secondary spectrum as applied in
\citet{2014MNRAS.440L..36P}. 

\subsection{Possible improvements}

We discuss several strategies which can improve on the solution
accuracy.  The single biggest improvement would be to monitor the speckle pattern over
several months, as the pulsar crosses each individual lens,
including both lensing systems.  This allows a direct comparison of
single lens to double-lens arclets.

Angular resolution can be improved using longer baselines, for example
adding a GMRT-GBT baseline doubles the resolution.  Observing at
multiple frequencies over a longer period allows for a more precise
measurement: when the pulsar is between two lenses, the refraction
angle $\beta$ is small, and one expects to see the lensing at higher
frequency, where the resolution is higher, and distances between
lens positions can be measured to much higher accuracy.

Holographic techniques \citep{2008MNRAS.388.1214W,2014MNRAS.440L..36P}
may be able to measure delays, fringe rates, and VLBI positions
substantially more accurately.  Combining these techniques, the
interstellar lensing could conceivably achieve distance measurements
an order of magnitude better than the current published effective
distance errors.  This could bring most pulsar timing array targets
into the coherent timing regime, enabling arc minute localization of
gravitational wave sources, lifting any potential source confusion.

Ultimately, the precision of the lensing results would be limited by
the fidelity of the lensing model.  In the inclined sheet model, the
images move along fold caustics.  The straightness of these caustics
depends on the inclination angle, which in turn depends on the
amplitude of the surface waves.  This analysis concludes a high degree
of inclination, and thus high fidelity for geometric pulsar studies.

\section{Conclusions}
\label{sec:conclusions}
We have applied the inclined sheet model \citep{2014MNRAS.442.3338P} 
to archival apex data of PSR B0834+06.  The data is well-fit by two
linear lensing screens, with nearly plane-parallel geometry.  The
second screen provides a precision test with 10 observables (5 time delays and 5 differential frequencies) and 3 free
parameters (the marked points in Table \ref{table:double_lens_compare}).  The model fits the data to $\sim$ percent accuracy
on each of 7 data points.  
This natural consequence of very smooth
reconnection sheets is an unlikely outcome of ISM turbulence.  These
results, if extrapolated to multi-epoch observations of binary
systems, might result in accurate distance determinations and
opportunities for removing scattering induced timing errors.  This
approach also opens the window to measuring precise transverse motions
of the ionized ISM outside the Galactic plane.

\section{Acknowledgements}

We thank NSERC for support. We acknowledge helpful discussions with
Peter Goldreich and M. van Kerkwijk.  
We thank Michael Williams for photography help.  
Siqi Liu thanks Robert Main and JD Emberson for helpful discussions on
improving the expression of the content.  
The National Radio Astronomy Observatory is a facility of the National Science Foundation operated under cooperative agreement by Associated Universities, Inc.

\newcommand{\araa}{ARA\&A}   
\newcommand{\afz}{Afz}       
\newcommand{\aj}{AJ}         
\newcommand{\azh}{AZh}       
\newcommand{\aaa}{A\&A}      
\newcommand{\aas}{A\&AS}     
\newcommand{\aar}{A\&AR}     
\newcommand{\apj}{ApJ}       
\newcommand{\apjs}{ApJS}     
\newcommand{\apjl}{ApJ}      
\newcommand{\apss}{Ap\&SS}   
\newcommand{\baas}{BAAS}     
\newcommand{\jaa}{JA\&A}     
\newcommand{\mnras}{MNRAS}   
\newcommand{\nat}{Nat}       
\newcommand{\pasj}{PASJ}     
\newcommand{\pasp}{PASP}     
\newcommand{\paspc}{PASPC}   
\newcommand{\qjras}{QJRAS}   
\newcommand{\sci}{Sci}       
\newcommand{\solphys}{Solar Physics}       %
\newcommand{\sova}{SvA}      
\newcommand{\aap}{A\&A}
\newcommand\jcap{{J. Cosmology Astropart. Phys.}}%
\newcommand{\prd}{Phys. Rev. D}

\bibliography{distance}

\begin{thebibliography}{}

\bibitem[\protect\citeauthoryear{Arnold}{Arnold}{1990}]{Arnold1990}
Arnold V.~I.,  1990, Singularities of Caustics and Wave Fronts.
Springer Netherlands

\bibitem[\protect\citeauthoryear{{Boyle} \& {Pen}}{{Boyle} \&
  {Pen}}{2012}]{boyle2012}
{Boyle} L.,  {Pen} U.-L.,  2012, \prd, 86, 124028

\bibitem[\protect\citeauthoryear{{Braithwaite}}{{Braithwaite}}{2015}]{2015MNRAS.450.3201B}
{Braithwaite} J.,  2015, \mnras, 450, 3201

\bibitem[\protect\citeauthoryear{{Brisken}, {Macquart}, {Gao}, {Rickett},
  {Coles}, {Deller}, {Tingay} \& {West}}{{Brisken}
  et~al.}{2010}]{2010ApJ...708..232B}
{Brisken} W.~F.,  {Macquart} J.-P.,  {Gao} J.~J.,  {Rickett} B.~J.,  {Coles}
  W.~A.,  {Deller} A.~T.,  {Tingay} S.~J.,    {West} C.~J.,  2010, \apj, 708,
  232

\bibitem[\protect\citeauthoryear{{Clegg}, {Fey} \& {Lazio}}{{Clegg}
  et~al.}{1998}]{1998ApJ...496..253C}
{Clegg} A.~W.,  {Fey} A.~L.,    {Lazio} T.~J.~W.,  1998, \apj, 496, 253

\bibitem[\protect\citeauthoryear{{Deller}, {Archibald}, {Brisken},
  {Chatterjee}, {Janssen}, {Kaspi}, {Lorimer}, {Lyne}, {McLaughlin}, {Ransom},
  {Stairs} \& {Stappers}}{{Deller} et~al.}{2012}]{deller12b}
{Deller} A.~T.,  {Archibald} A.~M.,  {Brisken} W.~F.,  {Chatterjee} S.,
  {Janssen} G.~H.,  {Kaspi} V.~M.,  {Lorimer} D.,  {Lyne} A.~G.,  {McLaughlin}
  M.~A.,  {Ransom} S.,  {Stairs} I.~H.,    {Stappers} B.,  2012, \apjl, 756,
  L25

\bibitem[\protect\citeauthoryear{{Deller}, {Boyles}, {Lorimer}, {Kaspi},
  {McLaughlin}, {Ransom}, {Stairs} \& {Stovall}}{{Deller}
  et~al.}{2013}]{deller13a}
{Deller} A.~T.,  {Boyles} J.,  {Lorimer} D.~R.,  {Kaspi} V.~M.,  {McLaughlin}
  M.~A.,  {Ransom} S.,  {Stairs} I.~H.,    {Stovall} K.,  2013, \apj, 770, 145

\bibitem[\protect\citeauthoryear{{Efron} \& {Tibshirani}}{{Efron} \&
  {Tibshirani}}{1991}]{efron91a}
{Efron} B.,  {Tibshirani} R.,  1991, Science, 253, 390

\bibitem[\protect\citeauthoryear{{Goldreich} \& {Sridhar}}{{Goldreich} \&
  {Sridhar}}{2006}]{2006ApJ...640L.159G}
{Goldreich} P.,  {Sridhar} S.,  2006, \apjl, 640, L159

\bibitem[\protect\citeauthoryear{{Kramer}, {Stairs}, {Manchester},
  {McLaughlin}, {Lyne}, {Ferdman}, {Burgay}, {Lorimer}, {Possenti}, {D'Amico},
  {Sarkissian}, {Hobbs}, {Reynolds}, {Freire} \& {Camilo}}{{Kramer}
  et~al.}{2006}]{2006Sci...314...97K}
{Kramer} M.,  {Stairs} I.~H.,  {Manchester} R.~N.,  {McLaughlin} M.~A.,  {Lyne}
  A.~G.,  {Ferdman} R.~D.,  {Burgay} M.,  {Lorimer} D.~R.,  {Possenti} A.,
  {D'Amico} N.,  {Sarkissian} J.~M.,  {Hobbs} G.~B.,  {Reynolds} J.~E.,
  {Freire} P.~C.~C.,    {Camilo} F.,  2006, Science, 314, 97

\bibitem[\protect\citeauthoryear{Longuet-Higgins}{Longuet-Higgins}{1960}]{LonguetHiggins1960}
Longuet-Higgins M.~S.,  1960, J. Opt. Soc. Am., 50, 845

\bibitem[\protect\citeauthoryear{{Lorimer} \& {Kramer}}{{Lorimer} \&
  {Kramer}}{2012}]{2012hpa..book.....L}
{Lorimer} D.~R.,  {Kramer} M.,  2012, {Handbook of Pulsar Astronomy}.
Cambridge University Press

\bibitem[\protect\citeauthoryear{{Pen} \& {King}}{{Pen} \&
  {King}}{2012}]{2012MNRAS.421L.132P}
{Pen} U.-L.,  {King} L.,  2012, \mnras, 421, L132

\bibitem[\protect\citeauthoryear{{Pen} \& {Levin}}{{Pen} \&
  {Levin}}{2014}]{2014MNRAS.442.3338P}
{Pen} U.-L.,  {Levin} Y.,  2014, \mnras, 442, 3338

\bibitem[\protect\citeauthoryear{{Pen}, {Macquart}, {Deller} \&
  {Brisken}}{{Pen} et~al.}{2014}]{2014MNRAS.440L..36P}
{Pen} U.-L.,  {Macquart} J.-P.,  {Deller} A.~T.,    {Brisken} W.,  2014,
  \mnras, 440, L36

\bibitem[\protect\citeauthoryear{{P{\'e}rez-Garc{\'{\i}}a}, {Silk} \&
  {Pen}}{{P{\'e}rez-Garc{\'{\i}}a} et~al.}{2013}]{2013PhLB..727..357P}
{P{\'e}rez-Garc{\'{\i}}a} M.~{\'A}.,  {Silk} J.,    {Pen} U.-L.,  2013, Physics
  Letters B, 727, 357

\bibitem[\protect\citeauthoryear{{Shapiro} \& {Field}}{{Shapiro} \&
  {Field}}{1976}]{1976ApJ...205..762S}
{Shapiro} P.~R.,  {Field} G.~B.,  1976, \apj, 205, 762

\bibitem[\protect\citeauthoryear{{Stinebring}, {McLaughlin}, {Cordes},
  {Becker}, {Goodman}, {Kramer}, {Sheckard} \& {Smith}}{{Stinebring}
  et~al.}{2001}]{2001ApJ...549L..97S}
{Stinebring} D.~R.,  {McLaughlin} M.~A.,  {Cordes} J.~M.,  {Becker} K.~M.,
  {Goodman} J.~E.~E.,  {Kramer} M.~A.,  {Sheckard} J.~L.,    {Smith} C.~T.,
  2001, \apjl, 549, L97

\bibitem[\protect\citeauthoryear{{Walker} \& {Wardle}}{{Walker} \&
  {Wardle}}{1998}]{1998ApJ...498L.125W}
{Walker} M.,  {Wardle} M.,  1998, \apjl, 498, L125

\bibitem[\protect\citeauthoryear{{Walker}, {Koopmans}, {Stinebring} \& {van
  Straten}}{{Walker} et~al.}{2008}]{2008MNRAS.388.1214W}
{Walker} M.~A.,  {Koopmans} L.~V.~E.,  {Stinebring} D.~R.,    {van Straten} W.,
   2008, \mnras, 388, 1214

\end{thebibliography}
\bibliographystyle{mn2e}

\label{lastpage}

\end{document}